\documentclass{ws-ijmpd}
\usepackage[super,compress]{cite}
\begin{document}

\markboth{Oleg Yu. Tsupko and Gennady S. Bisnovatyi-Kogan}
{Black Hole Shadow in McVittie Metric}

%
\catchline{}{}{}{}{}
%

\title{FIRST ANALYTICAL CALCULATION\\ OF BLACK HOLE SHADOW IN MCVITTIE METRIC}

\author{OLEG YU. TSUPKO}

\address{Space Research Institute of Russian Academy of Sciences,\\Profsoyuznaya 84/32, Moscow 117997, Russia\\
tsupko@iki.rssi.ru; tsupkooleg@gmail.com}

\author{GENNADY S. BISNOVATYI-KOGAN}

\address{Space Research Institute of Russian Academy of Sciences,\\
Profsoyuznaya 84/32, Moscow 117997, Russia\\
National Research Nuclear University MEPhI (Moscow Engineering Physics Institute),\\
Kashirskoe Shosse 31, Moscow 115409, Russia\\
Moscow Institute of Physics and Technology,\\
9 Institutskiy per., Dolgoprudny, Moscow Region, 141701, Russia\\
gkogan@iki.rssi.ru}

\maketitle

\begin{history}
\received{Day Month Year}
\revised{Day Month Year}
\end{history}

\begin{abstract}
Cosmic expansion influences the angular size of black hole shadow. The most general way to describe a black hole embedded into an expanding universe is to use the McVittie metric. So far, the exact analytical solution for the shadow size in the McVittie metric, valid for arbitrary law of expansion and arbitrary position of the observer, has not been found. In this paper, we present the first analytical solution for angular size of black hole shadow in McVittie metric as seen by observer comoving with the cosmic expansion. We use a method of matched asymptotic expansions to find approximate solution valid within the entire range of possible positions of observer. As two particular examples, we consider black hole in de Sitter and matter dominated universe.
\end{abstract}

\keywords{black hole shadow; McVittie metric; cosmological expansion}

\ccode{PACS numbers:}


\section{Introduction}

The shadow of a black hole is a dark spot in the sky in the direction of the black hole, in  presence of the background of other light sources. The shadow is formed due to strong deflection of light rays in vicinity of the black hole and absorption of light rays by the black hole. After recent observational evidence \cite{EHT-1, EHT-2, EHT-3, EHT-4, EHT-5, EHT-6}, great attention is now focused on various aspects of the study of the shadow of black holes \cite{Synge-1966, Bardeen-1973, Luminet-1979, Chandra, Dymnikova-1986, Falcke-2000, Takahashi-2004, HiokiMaeda2009, Moscibrodzka-2009,  Psaltis-2010, Frolov-Zelnikov-2011, Bambi2013, Tsukamoto-2014, Perlick-2014, Zakharov-2014, Lu-Broderick-2014, Perlick-2015, Interstellar-2015, Cunha-PRL-2015, Perlick-Tsupko-BK-2015, Rezzolla-Ahmedov-2015, Johannsen-2016, Konoplya-2016a, Shipley-2016, Perlick-Tsupko-2017, Tsupko-2017, black-hole-cam-2017, cunha-2017, Doeleman-2017, Eiroa-2018, Tsukamoto-2018, Cunha-2018, Mizuno-NatAstr-2018, Mars-2018, Yunes-2018, Kumar-2018, Gyulchev-2018, Singh-Ghosh-2018, Yunes-2019, Dokuchaev-2019, Gralla-2019, Johnson-2019, Narayan-2019, Konoplya-2019, Dokuchaev-2019-review, Jusufi-2019,  Shaikh-2019, Ishihara-2019, Alexeyev-2019, Mann-2019}.

Since we live in the expanding universe, this expansion
should influence the size of a black hole shadow. In order to calculate the angular size of the BH shadow in an expanding universe, it is necessary to take into account both the gravity of the black hole and  cosmological expansion at each part of the light trajectory. Moreover, the trajectories have to be calculated in strong gravity regime. All this makes the calculations non-trivial, and an exact analytical solution for the shadow size valid for an arbitrary position of the observer in the general case of the Friedmann universe has not yet been found.

So far, the exact analytical solution for black hole shadow in expanding universe is found only for particular case: when the expansion is driven by cosmological constant only. To model Schwarzschild black hole embedded in a de Sitter universe, Kottler \cite{Kottler1918} (also known as the Schwarzschild-de Sitter) spacetime can be used. Black hole shadow in static representation of Kottler metric as seen by static observer was found in paper of Stuchl\'{i}k and Hled\'{i}k \cite{Stuchlik-1999}, see also Refs.~\citen{Stuchlik-2007, Stuchlik-2018}. The first calculation of the shadow angular size as seen by observer comoving with the cosmic expansion has been performed by Perlick, Tsupko and Bisnovatyi-Kogan \cite{Perlick-Tsupko-BK-2018}.

Recently, it has been shown \cite{BK-Tsupko-2018} that angular size redshift relation (an apparent angular size written via the angular diameter distance) can serve as a good approximate solution for the shadow size in the general case of expanding FRW universe. It can be done under two realistic assumptions, that observer is far from black hole, and that the cosmic expansion is negligible near black hole. If such conditions hold, influence of a black hole gravity on the light propagation in the expanding universe remains only in finding the value
of the black hole shadow effective linear size, for using it in the formula for the angular size. Basing on this result, we suggested to use a black hole shadow as a standard ruler for finding cosmological parameters \cite{ruler}. Later it was discussed in Refs.~\citen{Qi-Zhang-2019, Vagnozzi-2020}.

The most general way to describe a black hole embedded in the expanding universe is to use the McVittie metric found in 1933 \cite{McVittie-1933}. Properties of this metric including the equations of motion have been studied in series of papers of Nolan \cite{Nolan-1, Nolan-2, Nolan-3}. Subsequently, McVittie spacetime has been studied in papers of Kaloper, Kleban and Martin \cite{Kaloper-2010}; Carrera and Giulini \cite{Carrera-Giulini-2010a}; Lake and Abdelqader \cite{Lake-Abdelqader-2011}; Anderson \cite{Anderson-2011}; Nandra, Lasenby and Hobson \cite{Hobson-2012a, Hobson-2012b}; da Silva, Fontanini and Guariento \cite{Silva-2013}; Nolan \cite{Nolan-2014, Nolan-2017}. For review see the paper of Carrera and Giulini \cite{Carrera-Giulini-2010-review}. Gravitational lensing in McVittie in case of small deflection were examined in papers of Piattella \cite{Piattella-PRD-2016, Piattella-Universe-2016}; Aghili, Bolen and Bombelli \cite{Aghili-2017}; Faraoni and Lapierre-L\'{e}onard \cite{Faraoni-2017}.

The complexity of the light path calculation in McVittie metric is primarily associated with the explicit dependence of the metric on time and, as a consequence, in the absence of the 'energy' constant of motion. We recall that the Schwarzschild--de-Sitter metric can be rewritten in the static form, with both constants of the energy and angular momentum. But in McVittie case, only angular momentum is conserved. This leads to the fact that the equations of motion cannot be written in the form of differential equations of the first order. To calculate geodesics in McVittie, we need to solve second-order differential equations \cite{Aghili-2017, Carrera-Giulini-2010-review}, therefore it is difficult to proceed analytically. To the best of our knowledge, analytical results for light ray propagation in McVittie are obtained only for weak deflection situation \cite{Piattella-PRD-2016, Piattella-Universe-2016, Aghili-2017}. Since black hole shadow requires the calculation in strong gravity regime, it is not clear how to find the exact analytical solution for the shadow size in this metric. We note here that numerical calculation of shadow size using numerical integration of light geodesics in McVittie metric has been performed in Ref.~\refcite{BK-Tsupko-2018}.

In this paper we find the first analytical solution for the shadow size in McVittie metric valid for any expansion model and for any observer's position. Our idea is based on the fact, that typical parameters of the problem are such, that the method of matched asymptotic expansions \cite{match-01, match-02} may be used for its solution. The method is based on existence of two known asymptotic analytical solutions, and the overlap region where both solutions are valid. Matching two asymptotic solution allows not only to find some unknown constants, but also to construct the approximate composite solution valid for the whole interval of variable. This method was used in other areas of astrophysics. For example, it was applied for solving the problem of the structure of a boundary layer between the accretion disk and a neutron star \cite{regev,bk94}. Also, it was used for calculation of black hole metric in radiation-dominated universe \cite{dokuchaev-match}.

In McVittie case, two asymptotic solutions are the formula for shadow in Schwarzschild space-time and formula for angular size of object in FRW universe. As crucial feature, we also assume the existence of region of almost flat space-time between neighbourhood of the black hole and the regions of significant expansion. This condition is satisfied for all cosmological black holes. As a result, we write the composite solution valid everywhere: for any position of observer, including neighbourhood of black hole, distant regions and all intermediate regions.

The paper is organized as follows. In the next Section we describe the method in details and find the approximate analytical solutions for the shadow size. In Section 3 we consider two particular cases: black hole in de Sitter universe (Kottler metric) and black hole in the matter dominated Friedmann universe (Einstein de Sitter model). Then we go to Conclusions.

\section{Solution}

\subsection{General idea of method}

To apply the method, we need to have two asymptotic analytical solutions, usually called 'inner' and 'outer'. We also need to have an overlap region where both of these solutions should be valid. The existence of this region allows to match two solutions, finding unknown constants. Moreover, the method allows to write an approximate solution (usually called 'composite') that will be valid for the entire range of variable \cite{match-01, match-02}.

McVittie metric has a form 
\begin{equation} \label{McVittie}
ds^2 = - \left( \frac{1-\mu}{1+\mu} \right)^2 c^2 dt^2 + (1+\mu)^4 a^2(t) (d\rho^2 + \rho^2 d\Omega^2) ,
\end{equation}
\begin{equation}
\text{where} \;\; \mu = \frac{m}{2a(t) \rho}, \; d \Omega ^2 = \mathrm{sin} ^2 \vartheta \, d \varphi ^2 + d \vartheta ^2 \, ,
\end{equation}
$a(t)$ is the scale factor, $m = GM/c^2$ is mass parameter, with $M$ being the black hole mass. In asymptotic cases the McVittie metric is reduced to the Schwarzschild metric and to the Friedmann-Robertson-Walker (FRW) metric (see below).

We are interested in finding out what size of the shadow is observed by a comoving observer located at $\rho_O$ at the time $t_O$. As mentioned above, an exact analytic solution for angular size of BH shadow valid for arbitrary observer's position is not found yet.
Our goal is to find an approximate solution using the method described above. For that, we will match two asymptotic solutions, known for  Schwarzschild and FRW cases, $\alpha_{schw}$ and $\alpha_{cosm}$.

\subsection{Schwarzschild ('inner') solution}

With $a(t) \equiv \mbox{const}$, the McVittie metric (\ref{McVittie}) is simplified to the Schwarzschild metric in isotropic coordinates. Since we are interested in a non-trivial case $a(t) \not\equiv \mbox{const}$, we need to understand at what scales we can consider the scale factor as constant. Physically, expansion is negligible if the observer is close to a black hole, and redshift is small: $z \ll 1$. Indeed, we can write the scale factor $a(t)$ as \cite{Hobson}
\begin{equation}
a(t) = a(t_0) - (t_0-t) \, \dot{a} (t_0) + \, ... \, = a(t_0) \, [ 1 - (t_0-t) H(t_0)  + \, ...   \, ]   \,  .
\end{equation}
If $t_0-t \ll H_0^{-1}$, we may write that $a(t) \simeq a(t_0)$. For small values of $(t_0-t)$ we can write $z \simeq (t_0-t) H_0$ \cite{Hobson}, so we obtain finally:
\begin{equation}
a(t) \simeq \mbox{const} = a(t_0) \;\; \mbox{for} \;\; z \ll 1  \, .
\end{equation}

With $a(t)=a(t_0)$ the metric (\ref{McVittie}) is reduced to the form where all coefficients are independent on time:
\begin{equation} \label{McVittie-0}
ds^2 = - \left( \frac{1-\mu_0}{1+\mu_0} \right)^2 c^2 dt^2 + (1+\mu_0)^4 a^2(t_0) (d\rho^2 + \rho^2 d\Omega^2) ,
\end{equation}
\begin{equation}
\text{where} \;\; \mu_0 = \frac{m}{2a(t_0) \rho}.
\end{equation}

With the help of new variable
\begin{equation} \label{r-defin}
r = a(t_0) \,  \rho  \, ,
\end{equation}
time-independent metric (\ref{McVittie-0}) can be reduced to
the Schwarzschild metric in isotropic coordinates:
\begin{equation} \label{Schw-isotr}
ds^2 = - \left( \frac{1- m/2r}{1+m/2r} \right)^2 c^2 dt^2 + (1+m/2r)^2 (d r^2 + r^2 d\Omega^2) .
\end{equation}
Further, introducing the variable
\begin{equation} \label{R-defin}
R =  r \left( 1 + \frac{2m}{r} \right)^2    ,
\end{equation}
the usual form of Schwarzschild metric is recovered:
\begin{equation}  \label{Schw-usual}
ds^2 = - \left(1-\frac{2m}{R}\right) c^2 dt^2 + \frac{dR^2}{1-2m/R} \,  +  \, R^2 d\Omega^2  \,  .
\end{equation}
Note that $R=r$ if $m \to 0$.

Let us consider the comoving observer in the spacetime (\ref{McVittie-0}). An observer has a radial coordinate $\rho_O$ and observes the shadow at a time $t_O$. Obviously, the time of observation $t_O$ coincides with the present time $t_0$ introduced earlier. We denote the observer's position in new variables as
\begin{equation} \label{r-R-observer}
r_O = a(t_0) \rho_O \, , \quad  R_O =  r_O \left( 1 + \frac{2m}{r_O} \right)^2    .
\end{equation}

For the Schwarzschild metric in the form (\ref{Schw-usual}),
the angular size of the shadow is given by \cite{Synge-1966} as
\begin{equation} \label{synge}
\sin \alpha_{schw} = \frac{3\sqrt{3}m \sqrt{1 - 2m/R_O}}{R_O} \, .
\end{equation}
Noting that $0 < \alpha_{schw} \le \pi/2$ for $R_O \ge 3m$, and $\pi/2 \le \alpha_{schw} \le \pi$ for $2m \le R_O \le 3m$, we write the angle as
\begin{equation} \label{alpha-schw}
\alpha_{schw}(R_O) = \left\{\begin{array}{l}
 \pi - \mathrm{arcsin} \left( 3\sqrt{3}m \sqrt{1 - 2m/R_O}/R_O \right) \; \mbox{for}  \;\;   2m  \le R_O \le 3m \, ,\\
 \mathrm{arcsin} \left( 3\sqrt{3}m \sqrt{1 - 2m/R_O}/R_O \right) \; \mbox{for}  \;\;   R_O \ge 3m    \, .
\end{array} \right.
\end{equation}


\subsection{Cosmological ('outer') solution}

For $m/\rho \to 0$, that is at large distances from BH, the McVittie metric (\ref{McVittie}) is simplified to FRW metric written in comoving coordinates:
\begin{equation} \label{FRW}
ds^2 = - c^2 dt^2 +  a^2(t) (d \rho^2 + \rho^2 d\Omega^2) \, .
\end{equation}

Apparent angular size $\alpha$ of any object of known physical size $L$ as seen by comoving observer in expanding FRW universe can be calculated as \cite{Mattig-1958, Zeldovich-1964, Dashevsk-Zeldovich-1965, Zeldovich-Novikov-book-2, Hobson, Mukhanov-book}
\begin{equation} \label{ang-size}
\alpha = \frac{L}{D_A(z)}  \, ,
\end{equation}
where $D_A(z)$ is the angular diameter distance. In the flat $\Lambda$CDM model it can be written as \cite{Hobson, Jones-book}
\begin{equation}
D_A(z) = \frac{c}{1+z} \int \limits_0^z \frac{d\tilde{z}}{H(\tilde{z})} \, ,
\end{equation}
where
\begin{equation} \label{H-z-def}
H(\tilde{z}) = H_0 \left[ \Omega_{m0}(1+\tilde{z})^3 + \Omega_{r0} (1+\tilde{z})^4  + \Omega_{\Lambda0}  \right]^{1/2} \, .
\end{equation}
Here $H_0$ is the present value of the Hubble parameter $H(t_0)$, and $\Omega_{m0}$, $\Omega_{r0}$, $\Omega_{\Lambda 0}$ are the present dimensionless density parameters for matter, radiation and dark energy, respectively.

Since for $r \gg m$ the McVittie metric tends to FRW metric, we can use the formula (\ref{ang-size}) for calculation of angular size of the shadow for comoving observer at large distances:
\begin{equation} \label{cosm-sh}
\alpha_{cosm} = \frac{L_{\mathrm{sh}}}{D_A(z)} .
\end{equation}
The only crucial difference from FRW case without BH is that we don't know the value of $L_{\mathrm{sh}}$ which is now the effective linear radius of the shadow. Specific value $L_{\mathrm{sh}}$ is determined by strong bending of light rays in the vicinity of BH. We will find this unknown constant later, by matching two solutions. Note that formula (\ref{cosm-sh}) implies that the angle $\alpha_{cosm}$ is small.

Eq.(\ref{cosm-sh}) is written as a function of $z$, but we need to have it as a function of radial coordinate. For transformation, the following formula can be used \cite{Hobson, Jones-book}:
\begin{equation} \label{z-rO}
\rho_O a(t_0) = c \int \limits_0^z \frac{d\tilde{z}}{H(\tilde{z})} \, ,
\end{equation}
where $H(\tilde{z})$ is defined in eq.(\ref{H-z-def}). Further, it is more convenient to use $R_O$ instead of $r_O$. At large distances from the black hole, $r \gg m$, the variables of $r$ and $R$ coincide. Therefore we can write the connection between $R_O$ and $z$:
\begin{equation} \label{z-RO}
R_O = c \int \limits_0^z \frac{d\tilde{z}}{H(\tilde{z})} \, .
\end{equation}

Substitution of $z$ expressed via $R_O$ into (\ref{cosm-sh}) gives the $\alpha_{cosm}$ as a function of $R_O$.

\subsection{Overlap region}

Now we have two solutions, $\alpha_{schw}(R_O)$ and $\alpha_{cosm}(R_O)$, which we would like to match.

Idea of matching is based on existence of overlap region where both solutions should be valid. To connect the expansions, we introduce an intermediate variable which is to be located within the overlap region \cite{match-01}. Intermediate variable $R_{int}$ should satisfy the condition:
\begin{equation} \label{R-int-0}
R_{inner} \ll R_{int} \ll R_{outer} \, ,
\end{equation}
where $R_{inner}$ and $R_{outer}$ are characteristic scales of inner and outer solutions validity. To match the expansions, the inner and outer approximations must give the same result when the intermediate value is substituted.

In our problem, the overlap region is an almost flat space-time between the black hole and the region of rapid expansion. On the one hand, this region is far enough from the black hole so that its gravity can be considered as negligible. On the other hand, the scales are still small enough so that cosmic expansion can be neglected.

First condition is $R \gg m$. Second condition can be obtained from the condition $z \ll 1$. With using of relation (\ref{z-RO}) for small $z$, this leads to $R \ll c/H_0$. Finally, we obtain the condition for overlap region and intermediate variable as
\begin{equation} \label{R-int-1}
m \ll R_{int} \ll \frac{c}{H_0} \, .
\end{equation}
To conclude: for given $m$ and $H_0$, the matching is possible only if it is possible to introduce the value $R_{int}$ which satisfy the condition (\ref{R-int-1}). Similar condition was used in our previous paper to find the solution at large cosmological distances \cite{BK-Tsupko-2018}.

\subsection{Matching solutions}

For $R_O \gg m$, the solution (\ref{alpha-schw}) is simplified to
\begin{equation} \label{alpha-pract-1}
\alpha_{schw}^{int}(R_O) = \frac{3\sqrt{3} m }{R_O} \, .
\end{equation}
At the same time, for $R_O \ll c/H_0$ ($z \ll 1$), the angular diameter distance $D_A(z)$ is simplified to $r_O$ (which coincides with $R_O$ for $r \gg m$ ), and we obtain from eq. (\ref{cosm-sh}) that
\begin{equation} \label{cosm}
\alpha_{cosm}^{int}(R_O) =  \frac{L_{\mathrm{sh}}}{R_O} \, .
\end{equation}

Considering $R_O$ from the intermediate region (\ref{R-int-1}) 
we  match two solutions (\ref{alpha-pract-1}) and (\ref{cosm}) at this point, which
gives the value of constant $L_{\mathrm{sh}}$:
\begin{equation}
L_{\mathrm{sh}} = 3\sqrt{3}m .
\end{equation}
Substituting $L_{\mathrm{sh}}$ into formula (\ref{cosm-sh}), we find that the outer asymptotic solution equals to:
\begin{equation} \label{cosm-sh-final}
\alpha_{cosm} = \frac{3\sqrt{3}m}{D_A(z)} .
\end{equation}
Solution (\ref{cosm-sh-final}) was constructed in our previous paper \cite{BK-Tsupko-2018} by using physical reasons.

\subsection{Composite solution}

A composite solution is an approximation valid for arbitrary value of the variable $R_O$. It is constructed as \cite{match-01}
\begin{equation} \label{alpha-compos}
\alpha_{appr}(R_O) =  \alpha_{schw}(R_O) +  \alpha_{cosm}(R_O)   -  \alpha_{overlap}(R_O)  \, .
\end{equation}
Here the overlap value is equal to the inner solution (\ref{alpha-pract-1}) or to the outer solution (\ref{cosm}) in the intermediate region. So, for $R_O$ inside the intermediate region $R_{int}$ we obtain
\begin{equation}
\alpha_{overlap}(R_O) = \alpha_{schw}^{int}(R_O) = \alpha_{cosm}^{int}(R_O) = \frac{3\sqrt{3}m }{R_O}  \, .
\end{equation}
Finally, we have the resulting composite solution defined by Eq.(\ref{alpha-compos}) where $\alpha_{schw}(R_O)$ is defined in (\ref{alpha-schw}), $\alpha_{cosm}(R_O)$ is defined in (\ref{cosm-sh-final}), and $z$ is expressed through $R_O$ by formula (\ref{z-RO}).


\section{Particular cases}

\subsection{Black hole in de Sitter universe}

\begin{figure}
\begin{center}
\includegraphics[width=0.9\textwidth]{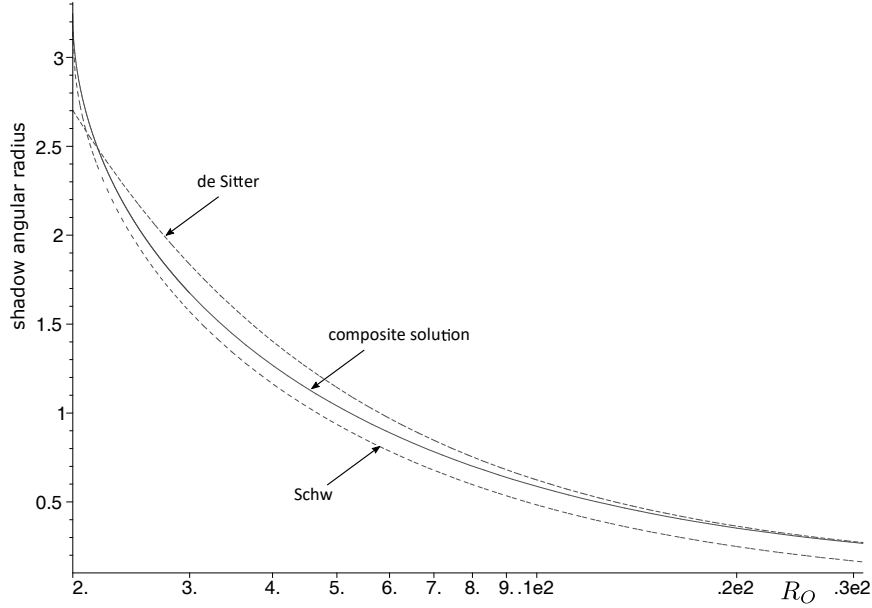}
\end{center}
\caption{Shadow angular radius $\alpha_{appr}$ in units of $mH_0/c$ as a function of $R_O$ in units of $m$ for black hole in de Sitter universe (solid line), see eq. (\ref{kottler-appr}). The Hubble value equals to $mH_0/c=0.02$. Inner solution $\alpha_{schw}$ and outer solution $\alpha_{cosm}$ are also shown by dashed lines.}
\label{fig:lambda}
\end{figure}

We start from the black hole embedded into de Sitter universe. This can be described by Kottler (Schwarzschild-de-Sitter) solution, which is a particular case of the McVittie metric. For this case we have:
\begin{equation}
\Omega_{\Lambda 0}= 1, \; H(t) = H_0 = \mbox{const}, \; a(t) = e^{H_0 t} .
\end{equation}
The formula (\ref{cosm-sh-final}) gives that
\begin{equation} \label{kottler-cosm}
\alpha_{cosm} = \frac{3\sqrt{3}m H_0}{c} \, \frac{1+z}{z} \, .
\end{equation}
The formula (\ref{z-RO}) gives
\begin{equation} \label{kottler-z-RO}
z = \frac{H_0 R_O}{c} \, .
\end{equation}
After substitution of (\ref{kottler-z-RO}) into (\ref{kottler-cosm}), we obtain finally the approximate solution in the form:
\begin{equation} \label{kottler-appr}
\alpha_{appr}(R_O) = \alpha_{schw}(R_O) + \frac{3\sqrt{3}m H_0}{c}  \,  .
\end{equation}
Note that, due to condition (\ref{R-int-1}), the second term is small ($\ll 1$). The approximate solution is shown in Fig. ~\ref{fig:lambda}.

For the particular case of Kottler metric, the exact solution for angular size of the shadow is known. The shadow angular radius as seen by the comoving observer is defined by the expression \cite{Perlick-Tsupko-BK-2018}
\begin{equation} \nonumber
\sin \alpha_{exact}(R_O) =
\frac{\sqrt{27} \, m}{R_O} \sqrt{1-\frac{2m}{R_O}}
\sqrt{ 1 - \frac{27H_0^2m^2}{c^2}} \, \mp \,
\end{equation}
\begin{equation} \label{exact-kottler-0}
\, \mp \, \frac{\sqrt{27} \, m \, H_0}{c}
\sqrt{ 1 - \frac{27 m^2}{R_O^2} \left( 1 - \frac{2m}{R_O} \right)}
\, .
\end{equation}
Here the minus sign should be chosen for $R_{H1} < R_O < 3m$, and the plus sign for $3m < R_O < \infty$ ($R_{H1}$ is black hole event horizon size in Kottler metric). Using the formula for the sine of the sum, 
\begin{equation}
\sin (\alpha + \beta) = \sin \alpha \cos \beta + \cos \alpha \sin \beta    
\end{equation}
and properties of trigonometrical functions, we can rewrite (\ref{exact-kottler-0}) in the simple form:
\begin{equation} \label{exact-kottler-1}
\alpha_{exact}(R_O) = \alpha_{schw}(R_O) + \arcsin (\sqrt{27} m H_0/c) \, ,
\end{equation}
where $\alpha_{schw}$ is defined by Synge's formula, see (\ref{synge}) and (\ref{alpha-schw}).

The approximate solution (\ref{kottler-appr}) coincides with the exact one (\ref{exact-kottler-1}) because at small argument of arcsin we have $\arcsin x \approx x$.

\subsection{Black hole in matter dominated Friedmann  universe}

\begin{figure}
\begin{center}
\includegraphics[width=0.9\textwidth]{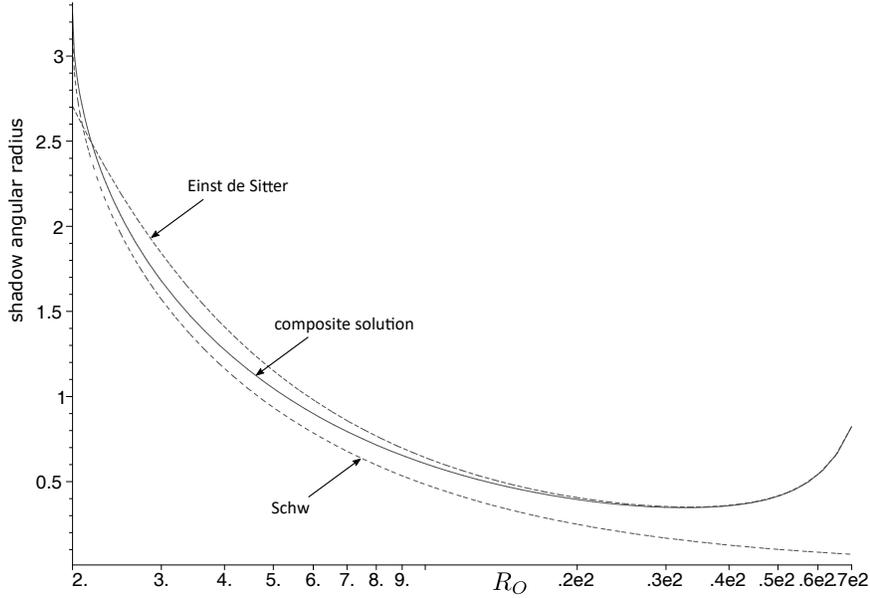}
\end{center}
\caption{Shadow angular radius $\alpha_{appr}$ in units of $mH_0/c$ as a function of $R_O$ in units of $m$ for black hole in matter dominated universe (solid line). The Hubble value equals to $mH_0/c=0.02$. Inner solution $\alpha_{schw}$ and outer solution $\alpha_{cosm}$ are also shown by dashed lines.}
\label{fig:matter}
\end{figure}

Let us consider the case of a black hole in the flat dusty Friedmann universe model (also known as Einstein-de-Sitter model) \cite{Hobson, Mukhanov-book}. This model has the following cosmological parameters: $\Omega_{m0} = 1$, $\Omega_{r0}=0$, $\Omega_{\Lambda0} =0$, $a(t) = (3H_0t/2)^{2/3}$.
For $\alpha_{cosm}$ we have:
\begin{equation}
\alpha_{cosm} (z) = 3\sqrt{3}m \, \frac{H_0}{2c} \, \frac{1+z}{[1-(1+z)^{-1/2}]} \, .
\end{equation}
The formula (\ref{z-RO}) gives
\begin{equation}
1 + z = \left( 1 - \frac{H_0 R_O}{2c} \right)^{-2}  \, .
\end{equation}
Finally, we obtain for $\alpha_{cosm}(R_O)$:
\begin{equation} \label{EdS-cosm}
\alpha_{cosm} (R_O) = 3\sqrt{3}m \, \frac{1}{R_O} \left( 1 - \frac{H_0 R_O}{2c} \right)^{-2} \, .
\end{equation}
Using (\ref{EdS-cosm}) in (\ref{alpha-compos}) gives the approximate solution  for $\alpha_{appr}$, which is plotted in Fig.~\ref{fig:matter}.

\section{Conclusions}

(i) We have found the approximate analytical solution for the angular size of the black hole shadow in McVittie spacetime (\ref{McVittie}) using the method of matched asymptotic expansions. Solution is valid for arbitrary position of observer $\rho_O$ at arbitrary time $t_O$. 

(ii) Variables $\rho_O$ and $t_O$ are included in the solution only as a parts of the following combination:
\begin{equation}
R_O =  a(t_O) \rho_O \left( 1 + \frac{2m}{a(t_O) \rho_O} \right)^2 \, .
\end{equation}
For the Schwarzchild metric, the variable $R_O$ coincides with the radial position of the observer in spherical coordinates. For FRW metric, $R_O$ coincides with proper distance. 

(iii) Method can be applied for the black hole in any expanding universe.

\section*{Acknowledgements}

This work is financially supported by the Russian Science
Foundation, Grant No. 18-12-00378.


\end{document}